\documentclass[sigconf]{acmart}

\renewcommand\footnotetextcopyrightpermission[1]{} 
\usepackage{multirow}
\usepackage{graphicx}
\usepackage{float}
\usepackage{booktabs}
\usepackage{amsmath} 

\AtBeginDocument{%
  }

\setcopyright{none}

\acmConference[]{}{}{} 
\acmYear{}
\acmPrice{}
\acmDOI{}
\acmISBN{}

\begin{document}

\title{Can GANs Learn the Stylized Facts of Financial Time Series?}


\author{Sohyeon Kwon}
\affiliation{%
  \institution{NCSOFT}
  \city{Seongnam}
  \country{Republic of Korea}}
\email{sohyeonk@ncsoft.com}

\author{Yongjae Lee}
\affiliation{%
  \institution{Ulsan National Institute of Science and Technology (UNIST)}
  \city{Ulsan}
  \country{Republic of Korea}}
\email{yongjaelee@unist.ac.kr}


\begin{abstract}
In the financial sector, a sophisticated financial time series simulator is essential for evaluating financial products and investment strategies. Traditional back-testing methods have mainly relied on historical data-driven approaches or mathematical model-driven approaches, such as various stochastic processes. However, in the current era of AI, data-driven approaches, where models learn the intrinsic characteristics of data directly, have emerged as promising techniques. Generative Adversarial Networks (GANs) have surfaced as promising generative models, capturing data distributions through adversarial learning. Financial time series, characterized “stylized facts” such as random walks, mean-reverting patterns, unexpected jumps, and time-varying volatility, present significant challenges for deep neural networks to learn their intrinsic characteristics. This study examines the ability of GANs to learn diverse and complex temporal patterns (i.e., stylized facts) of both univariate and multivariate financial time series. Our extensive experiments revealed that GANs can capture various stylized facts of financial time series, but their performance varies significantly depending on the choice of generator architecture. This suggests that naively applying GANs might not effectively capture the intricate characteristics inherent in financial time series, highlighting the importance of carefully considering and validating the modeling choices.
\end{abstract}




\maketitle

\section{Introduction}
The simulation of financial time series is crucial for designing investment strategies and financial products. Two approaches are commonly used in practice. A naïve approach involves testing strategies or products solely on historical data. However, as noted by \cite{Assefa-01}, this approach suffers from a lack of data. An alternative approach is to develop mathematical models of financial time series based on both academic research and practical observations of their properties. This model-based approach began with the use of simple Brownian motion to represent the unpredictable nature of stock prices. Over time, various models have been developed, including the Ornstein-Uhlenbeck (OU) process for mean-reverting behaviors, Jump Diffusion processes to account for occasional jumps, and the Heston model to capture time-varying volatilities. While these models are effective at reflecting known properties of financial time series, they can become highly complex when more realistic simulations are sought. Consequently, estimating the parameters of these sophisticated models becomes increasingly challenging.

Deep generative models have recently emerged as strong alternatives for simulating financial time series. Unlike model-driven approach, these models adopt a data-driven approach, learning various intrinsic characteristics of financial time series directly from the data without requiring specific modeling of certain properties. Among the various models, Generative Adversarial Networks (GANs) proposed by \cite{Goodfellow-01} have gained significant attention by employing an adversarial training mechanism between two distinct neural networks: a generator and a discriminator. GANs have demonstrated exceptional performance in image and text generation (e.g. \cite{Karras-01}; \cite{Ramesh-01}). Furthermore, there have been attempts to utilize GANs for time series generation, including financial time series (e.g., \cite{Yu-01}; \cite{yoon-01}). Notable examples of finance-specific GANs include QuantGANs \cite{wiese-01}, Tail-GAN \cite{cont-01} and Fin-GAN \cite{vuletic-01}.

Most previous studies have proposed GANs that generate log return distributions of financial time series, closely matching the shape of these distributions. \textbf{However, can financial time series generated by these GAN models be used to develop robust hedging strategies or investment models that cover both occasional and unanticipated situations and the sequence of such events?} Practically applicable GAN models must be able to mirror various patterns seen in real data, including random walks, mean reversion, jumps, and time-varying volatilities.

In this paper, we aim to verify the ability of GANs to learn various stylized facts inherent in financial time series: random walks, mean reversions, jumps, and time-varying volatilities. Each of these properties is represented by a specific stochastic process. For instance, the random walk property is modeled by Brownian motion or Geometric Brownian motion, while the Ornstein-Uhlenbeck process characterizes mean reversion. Jump diffusion models capture jump behaviors, and the Heston model depicts time-varying volatilities. Our goal is to evaluate whether GANs can accurately approximate these models, considering both return distributions and stylized facts. 

\section{Related Work}

\subsection{Model-driven financial time series simulator}

Traditionally, financial time series have been modeled using various stochastic processes, which are collections of random variables with a few parameters that distill the various characteristics of financial time series. Bachelier was the pioneer in modeling the prevalent randomness of financial time series by introducing the concept of Brownian motion. The Geometric Brownian motion, which simulates a lognormal distribution, models random walks with geometric growth and eliminates the possibility of stock prices being negative. Stock prices exhibit a tendency to revert to their mean, and their distributions are typically leptokurtic, deviating from the normal distribution. This behavior is best captured by the mean-reverting Ornstein-Uhlenbeck process, which is used in option pricing model (e.g., \cite{stein-01}). Real-world financial time series often exhibit heavy-tailed distributions, where extreme price movements occur more frequently than would be expected under a normal distribution, posing challenges for traditional models. In this regard, \cite{merton-01} developed the Jump Diffusion model, which incorporates the Poisson process for price jumps and successfully simulates both normal and abnormal events \cite{kou-01}. Additionally, financial markets are characterized by volatility clustering, where periods of high volatility are followed by high volatility and low volatility by low volatility, which reflects the time-varying nature of risk in financial markets.

\subsection{GANs in finance}

In recent years, the remarkable performance of GANs in various fields, including text and image generation, has spurred interest in their applicability within the financial sector. \cite{wiese-01} proposed QuantGAN, specifically tailored for financial time series generation, which uses a stacked temporal convolutional network to capture long-range temporal dependencies. \cite{koshiyama-01} used conditional GANs for calibrating trading strategies, integrating the generated samples into ensemble modeling. Meanwhile, \cite{pardo-01} trained a Wasserstein GAN with gradient penalty for data augmentation, demonstrating that synthetic data can mitigate overfitting and enhance trading strategies. \cite{cont-01} developed Tail-GAN by refining the loss function using the joint elicitability of tail risk information, such as value-at-risk and expected shortfall. The Fin-GAN, introduced by \cite{vuletic-01}, employs a novel economics-driven loss function to facilitate the forecasting of financial time series. On the other hand, \cite{kim2023gans} utilized the discriminator part of GANs to detect anomalies in financial time series, and \cite{kim2024enhancing} applied it to the mean-variance model to enable dynamic control of model robustness. For more examples of application of ML/AI models (including GANs) in finance, see \cite{lee2023overview}

\section{Models}

In this section, we examine the capability of GANs to capture the temporal structures of financial time series, as represented by the stochastic processes discussed in Section 2.1. While GANs do not require a specific architecture for the neural network, their performance varies depending on the chosen architecture. Additionally, GANs are known to be highly sensitive to hyperparameter configurations, particularly the learning rates for both the generator and discriminator. Therefore, the primary focus of our experiments is: (1) experimenting with various neural network architectures, and (2) formulating distribution distance-based objective functions for hyperparameter optimization.

\subsection{Model architectures}

GANs operate as a data generation framework via adversarial training and do not assume any specific network architecture. While the discriminator usually employs a simple architecture, the generator varies widely depending on the research objectives and the characteristics of the data. For instance, \cite{esteban-01} in RGAN and \cite{wiese-01} in QuantGAN, employed recurrent neural networks (RNN) and temporal convolutional networks (TCN) respectively, to encapsulate the temporal structure of time series data. Conversely, \cite{takahashi-01} experimented with multiple architectures, including multi-layer perceptrons (MLP), convolutional neural networks (CNN), and a hybrid of the two, MLP-CNN.

\begin{figure}[ht]
  \centering
  \includegraphics[width=\linewidth]{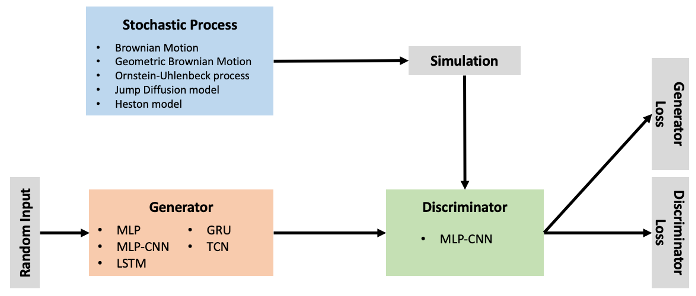}
  \caption{Experiment overview}
  \Description{Experiment overview}
\end{figure}

Therefore, we tested the following five different architectures for the generator: Multi-Layer Perceptron (MLP), Multi-Layer Perceptron – Convolutional Neural Network (MLP-CNN), Long Short-Term Memory (LSTM), Gated Recurrent Unit (GRU), and Temporal Convolutional Network (TCN). For simplicity, We fixed the architecture of the discriminator as MLP-CNN. An overview of the experiments in this study is shown in \textbf{Figure 1}. 

\subsection{Hyperparameter optimization}

GAN is extremely sensitive to hyperparameter settings, especially the learning rates of the generator and discriminator. Consequently, a hyperparameter search is essential to improve the model’s performance. In image generation, the quality of the generated images can be evaluated using the bias and variance of Fréchet Inception Distance, which measures the similarity between distribution of real and generated image features. However, there is currently no widely accepted metric for evaluating the performance of GANs in generating time series, nor is there an established objective for minimizing hyperparameter optimization in this context.

To measure the distance between the real and generated distributions in the context of hyperparameter optimization, we can use a representative metric such as the Kullback-Leibler (KL) divergence. KL divergence is a common measure of the difference between two probability distributions and can be employed as an objective function for hyperparameter optimization in GAN-based time series generation. This approach allows us to quantify how well the generated data approximates the real data, thereby guiding the optimization process to improve the quality of the generated time series.
\begin{math}
  D_{KL}(P||Q)
\end{math}
measures the average information entropy required for approximating distribution Q to model the actual distribution P and is defined as

\begin{equation}
  D_{KL}(P||Q) = - \int P(x) \ln{\{\frac{q(x)}{p(x)}}\}dx
\end{equation}

However, KL-divergence cannot measure how far the two probability distributions are due to its asymmetricity. Alternatively, a symmetrized and smoothed version of KL-divergence, Jensen-Shannon divergence, can be used to calculate the distance between the two distributions. Jensen-Shannon divergence lets \textit{M} be an average distribution of \textit{P} and \textit{D}, and it quantifies the average of 
\begin{math}
    D_{KL}(P||M) 
\end{math} and 
\begin{math}
    D_{KL}(Q||M)
\end{math}.

\begin{equation}
    JSD(P||Q) = \frac{1}{2}D_{KL}(P||M) + \frac{1}{2}D_{KL}(Q||M)
\end{equation}
\begin{displaymath}
    \textit{where } M = \frac{1}{2}(P + Q)
\end{displaymath}

To define the objective function for hyperparameter search, we use a linear combination of the Jensen-Shannon divergence between the log return distributions of the real and generated data, and the divergence between their final value distribution. The log return distribution reflects the volatility of the financial time series, which is a crucial feature when evaluating a financial time series simulator. Including the final value distribution in the objective function helps to prevent model collapse and retain the scale range of the generated data. Empirical weights are assigned to the log return and final value distributions to balance their contributions to the objective function effectively. Let \textit{R} and \textit{F} be the distribution of log return and final value, respectively, and the objective function is 

\begin{equation}
    min(4 * JSD(R_{real}||R_{generated})) + (1 * JSD(F_{real}||F_{generated}))
\end{equation}

In the case of multivariate time series, it is essential to capture not only the return and final value distributions but also the dependencies between variables during GAN training. Accurately modeling these dependencies ensures that the generated time series reflects the complex interactions present in the real data, which is crucial for applications such as portfolio optimization and risk management. To address this, the objective function for hyperparameter search should include metrics that assess the quality of these dependencies, ensuring that the generated data maintains the inherent correlations and relationships between variables. Let 
\begin{math}
    N_{0}
\end{math} and
\begin{math}
    N_{1}
\end{math} be the independent two normal distributions with the same dimension \textit{k}, then KL-divergence for multivariate normal distributions is defined as

\begin{equation}
    D_{KL}(N_{0}||N_{1})=
\end{equation}
\begin{displaymath}
    \frac{1}{2}(tr(\Sigma_{1}^{-1}\Sigma_{0}))-k + (\mu_{1}-\mu_{0})^{T}\Sigma_{1}^{-1}(\mu_{1}-\mu_{0})+\ln{(\frac{det\Sigma_{1}}{det\Sigma_{0}})}
\end{displaymath}

Including the marginal and bivariate distribution information, the hyperparameter is searched by the following function. 

\begin{equation}
    min(2 * D_{KLmultivariate})
\end{equation}
\begin{displaymath}
    + (4*(JSD(R_{realx1}||R_{generatedx1})+JSD(R_{realx2}||R_{generatedx2})))
\end{displaymath}
\begin{displaymath}
    + JSD(F_{realx1}||F_{generatedx1}) + JSD(F_{realx2}||F_{generatedx2})
\end{displaymath}

\section{Experiment}

\subsection{Data description}
For modeling temporal structures of financial time series, this study employs five stochastic processes as introduced in Section 2.1: Brownian motion (BM), Geometric Brownian motion (GBM), Ornstein-Uhlenbeck process (OU), Jump diffusion model (JD), and Heston model (HT). While historical real financial time series contain multiple inherent features, these features may not occur frequently, leading to a lack of data. However, with the availability of parameters, stochastic processes can be simulated consistently, providing a valuable tool for generating synthetic financial time series data that can capture the characteristics of the real-world data.

For explicit evaluation of the learning ability of GANs in financial time series, we generate each stochastic process with different parameter settings. For Brownian Motion, we vary the rate of change and the scale factor of the Wiener process. In the case of Geometric Brownian Motion (GBM), we adjust the drift and volatility parameters. For the Ornstein-Uhlenbeck (OU) process, the reverting speed and volatility are varied. The Jump Diffusion model is configured with different mean and standard deviation values for the normal distribution to describe jumps. Lastly, for the Heston (HT) model, we modify the volatility of volatility. Each stochastic process, except for the HT model, has six different parameter settings, while the HT model has three parameter settings. To assess the model’s performance based on sequence length, we create each stochastic process with three different sequence lengths: 25, 50, and 100. For the multivariate case, a bivariate Geometric Brownian motion is generated with two different correlation settings, high and low. The correlation between the variables is inherently generated within the multivariate normal distribution for the Wiener process of Geometric Brownian motion, with correlations set to approximately 0.9 and 0. The sequence length for the multivariate case is set to 50, and the drift and volatility for both variables (x1 and x2) are set to be different, 0.5 and 0.1, respectively, resulting in distinct return distribution shapes for each variable.

\begin{figure*}[ht]
  \centering
  \begin{tabular}{cc}
       \includegraphics[width=8cm, height=4cm]{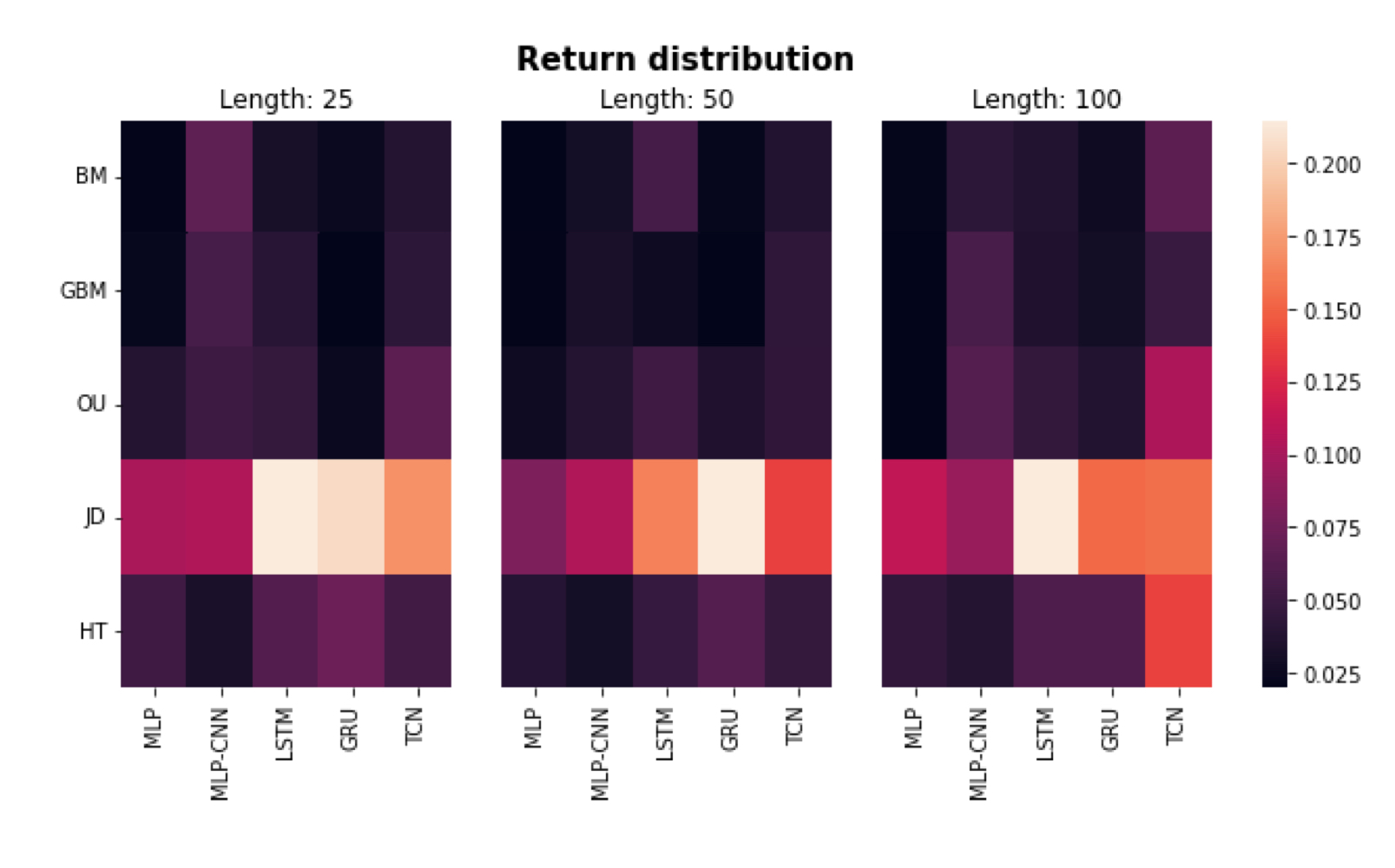}& \includegraphics[width=8cm, height=4cm]{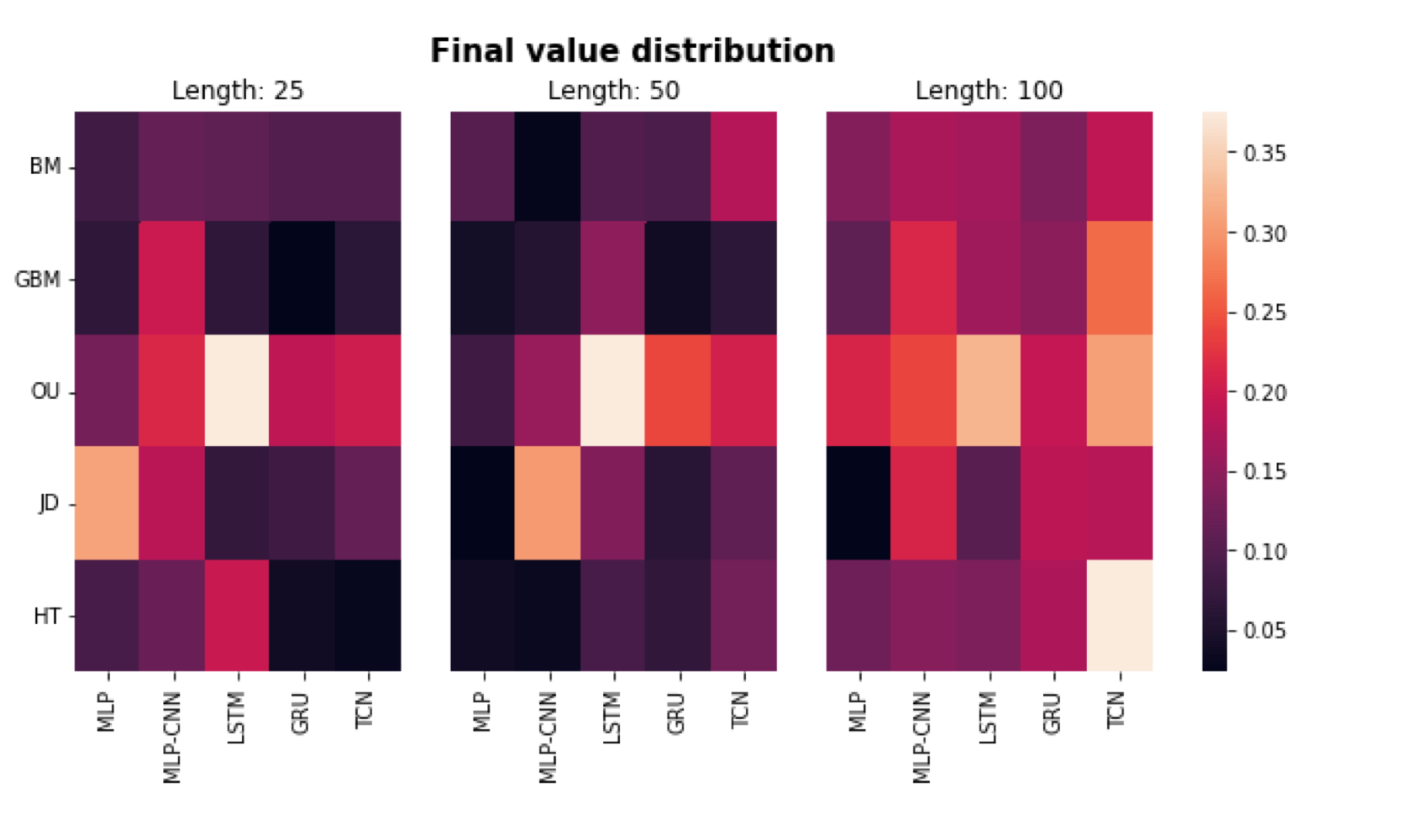} \\
  \end{tabular}
  \caption{Jensen-Shannon divergence heatmap of log return and final value distribution. Each value is the mean of every parameter type of each stochastic process.}
\end{figure*}

\begin{figure*}[ht]
  \centering
  \includegraphics[width=18cm, height=8cm]{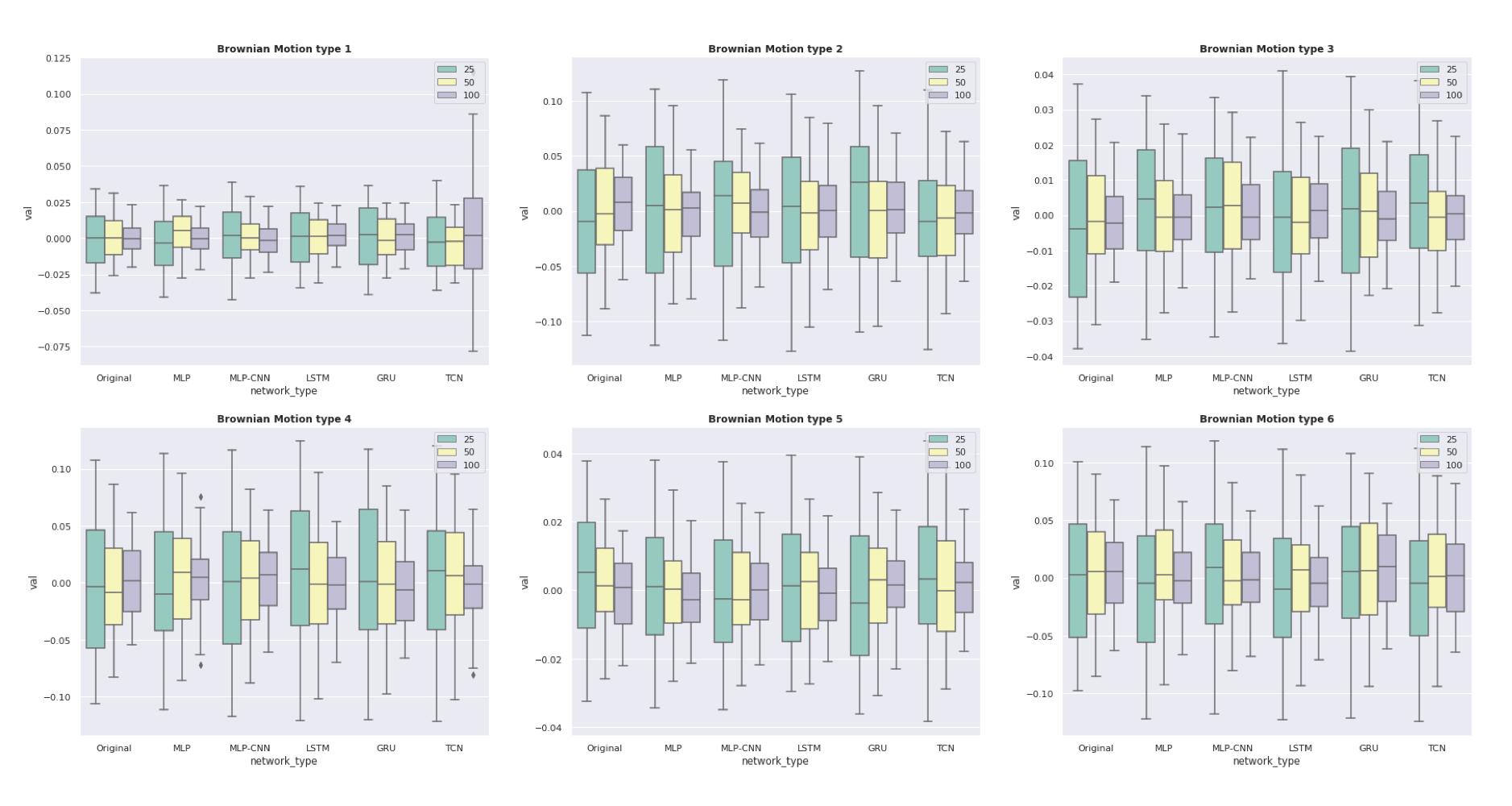}
  \caption{Box plots of 6 types of the Brownian motion. The first three boxes in each type frame refer to the statistical properties of the original case corresponding length. The following three boxplots are the statistical properties of synthetic data generated by each network. The types vary in terms of (mean, std) of Wiener process, with the parameters set to (0, 10), (0, 30), (2.5, 10), (2.5, 30), (5, 10), (5, 30).}
\end{figure*}

\subsection{Evaluation method}
We assess the quality of the synthesized financial time series from two perspectives: distribution evaluation and the characteristics of each stochastic process. Unfortunately, a widely accepted metric for evaluating the performance of time series generation models is unavailable. However, as Goodfellow et al., 2014 emphasize, GANs are successful in learning the underlying distribution of the training data. Therefore, we utilize the Jensen-Shannon divergence, as described in \textbf{Eq. 2}, as a distance measure between the distributions of the real and generated data. This metric is estimated for the overall evaluation of the model’s performance.

As the length of the Wiener process used as training data in this experiment is comparatively short, directly estimating the values of its parameters from the generated data is challenging. Therefore, we indirectly approximate the features of each stochastic from the generated data. For Brownian motion and Geometric Brownian motion, which both describe the randomness of the stock prices, we approximate their statistical properties, such as mean, standard deviation, maximum, and minimum of log return values. To approximate the reverting speed and count to the long-term mean of the Ornstein-Uhlenbeck process, we define the range of the long-term mean as $100 \pm margin$ and quantify the reverting speed by measuring the timesteps between two reverting events. We calculate the count as the number of reverting events in each time series. For the Jump Diffusion model, we estimate the size and count of jump events in each time series. Since classifying whether the value of each time step is in a jump or normal status is ambiguous, we define the value as a jump if $r_t \geq \mu_{\text{jump distribution}}- \sigma_{\text{jump distribution}}$ where $r_{t}$ refers to the return at time step $t$. Lastly for the Heston model, we estimate the volatility by rolling every five days and quantify its mean, standard deviation, maximum, and minimum values.

For the evaluation of multivariate time series, it is significant to learn the intrinsic characteristics of each variable and the dependencies between variables. The Jensen-Shannon divergence is estimated separately for each variable, x1 and x2, for independent variable evaluation. Variable dependency of multivariate time series is revealed in the bivariate distribution and correlation between variables. Therefore, the shape of the bivariate distribution is compared, and the correlations are quantified for comparison.

\begin{figure*}[ht]
  \centering
  \includegraphics[width=18cm, height=8cm]{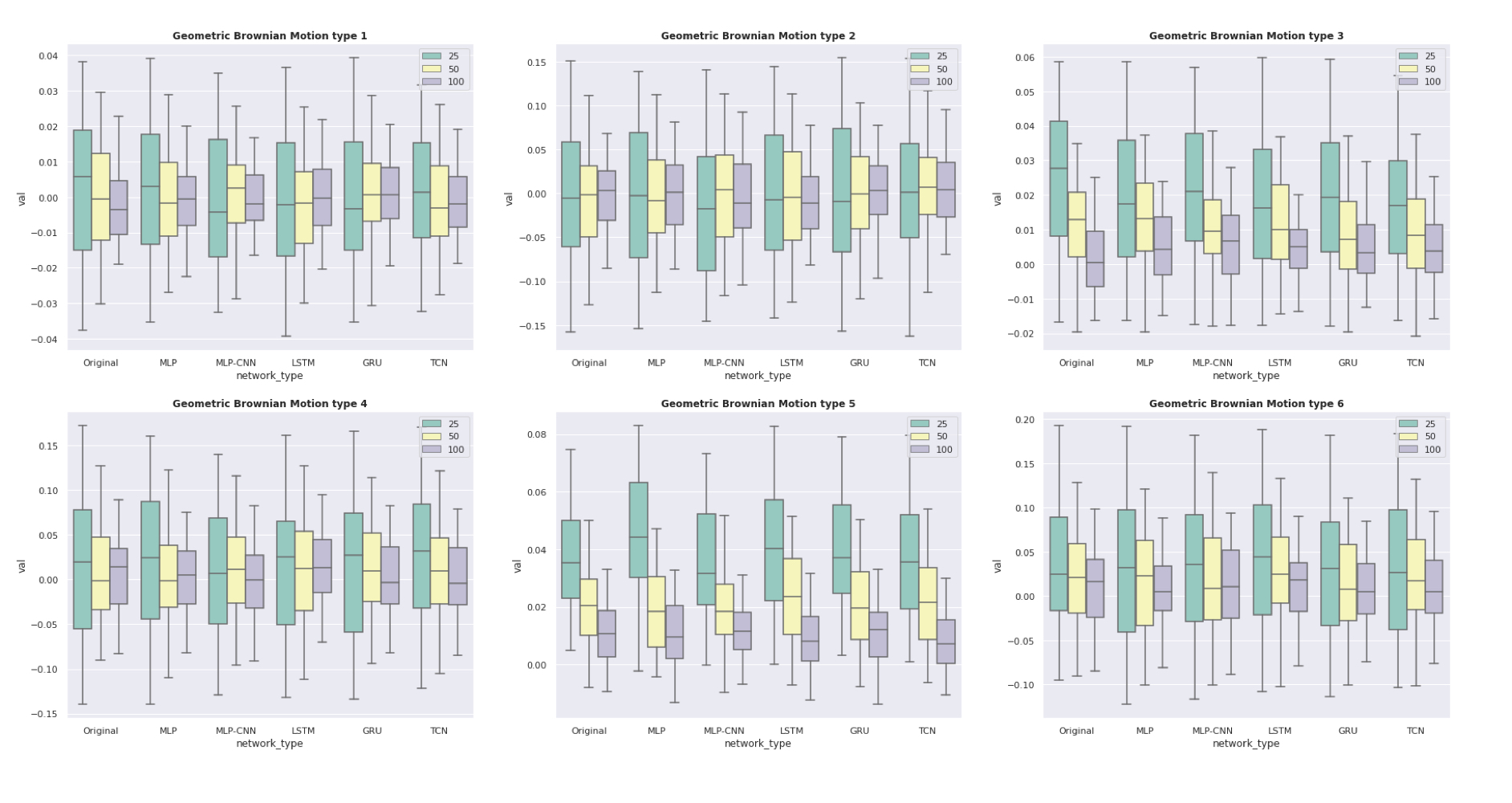}
  \caption{Box plots of 6 types of the Geometric Brownian motion. The first three boxes in each type frame refer to the statistical properties of the original case corresponding length. The following three boxplots are the statistical properties of synthetic data generated by each network. The types vary in terms of ($\mu$, $\sigma$) of Wiener process, with the parameters set to (0, 0.1), (0, 0.4), (0.5, 0.1), (0.5, 0.4), (1.0, 0.1), (1.0, 0.4).}
\end{figure*}

\subsection{Results}
\subsubsection{Univariate time series generation}

As GAN learns the distribution of the training data, the Jensen-Shannon divergence is used to estimate the distances between the log return and final value distributions, as shown in \textbf{Figure 2}. The length dependency is evident in the final value distribution, whereas there is little difference in return distribution between lengths. The Jensen-Shannon divergence of lengths 25 and 50 in the final value distribution of \textbf{Figure 2} generally remains low, but the values of length 100 are comparatively high. This indicates that as the length increases, there is a greater possibility of the generated final values being outside of the original data range, even though the return distributions resemble each other. This may occur because a small difference in return value between generated and original data results in a more significant difference as the error accumulates. Among the stochastic process types, Geometric Brownian motion shows the best learning performance both in log return distribution and final value distribution, with average values of 0.12913 and 0.46003, respectively, across the different lengths of the time series. As clearly shown in \textbf{Figure 2}, the log return distribution of generated data in the case of the Jump Diffusion model shows the most prominent distribution distance to that of the training data, with a value of 0.52243, regardless of the time series length. The final values of the Ornstein-Uhlenbeck process form a distribution that is the most dissimilar to the training data, quantified with an average distance value of 0.68728.

\begin{table*} 
  \label{tab:ouspeedcount}
  \begin{tabular}{cccccccccccccccccc}
    \toprule
     &&&& \multicolumn{2}{c}{Reverting Speed} &&&&& \multicolumn{2}{c}{Reverting Count} &&\\
    \midrule
    Length & Type & Original & MLP & MLP-CNN & LSTM & GRU & TCN & Original & MLP & MLP-CNN & LSTM & GRU & TCN \\
    \midrule
    \multirow{6}{*}{25} & 1 & 3.103	& 1.820	& 1.939	& \textbf{2.464}	& 1.943 & 1.835	& 4.850	& \textbf{3.892}	& 3.053	& 2.232	& 3.703	& 3.644\\
    & 2	& 3.481	& \textbf{3.409}	& 2.858	& 2.683	& 3.050	& 3.088	& 2.303 & 1.250 & 1.322 & \textbf{1.682} &	1.276 &	1.298 \\
    & 3	& 3.359	& \textbf{2.487}	& 2.069	& 1.767	& 1.757	& 2.058	& 5.530 & \textbf{3.794} &	3.414 &	3.177 &	3.680 &	3.001 \\
    & 4 & 4.145 & 3.514 &3.071 & 2.980 & \textbf{3.565} & 3.354 & 2.587 & 1.393 &	1.443 &	1.714 &	1.474 &	\textbf{1.729} \\
    & 5	& 3.423	& 2.241	& \textbf{2.269}	& 1.619	& 2.167	& 2.155	& 6.225 & 3.624 &	2.758 &	\textbf{5.112} &	3.628 &	3.537 \\
    & 6	& 4.662	& \textbf{4.420}	& 3.829	& 4.188	& 3.968	& 13.882 &2.982 & 1.782 &	\textbf{1.802} &	1.033 & 1.522 &1.583 \\
    \midrule
    \multirow{6}{*}{100} & 1 & 4.301 & 2.326 & \textbf{2.443} & 2.289 & 2.019 & 2.170	& 17.894 & 12.670 &	12.410 & 6.602 & \textbf{14.405}	& 13.891 \\
    & 2	& 8.830	& 5.565	& 4.912	& \textbf{7.088}	& 4.839	& 4.374 & 6.752 & 5.372 &	5.186 &	3.601 &	5.980 &	\textbf{6.617} \\
    & 3	& 4.395	& 2.909	& 2.721	& \textbf{3.056}	& 1.859	& 2.819	& 20.971 & 15.481 & 11.590 & 4.858 & \textbf{19.259} & 14.424 \\
    & 4	& 9.750	& \textbf{6.400}	& 5.069	& 5.420	& 4.539	& 4.053	& 8.049 & 5.702 &	4.996 &	5.672 &	7.170 &	\textbf{7.471} \\
    & 5	& 4.185	& 2.105	& 2.257	& \textbf{3.099}	& 2.142	& 2.006	& 24.076 & 14.875 & \textbf{14.930} & 4.822	& 13.874 & 9.387 \\
    & 6	& 9.474	& \textbf{5.418}	& 4.915	& 4.944	& 4.850	& 5.291	& 9.831 & 6.548 &	6.670 &	8.772 &	\textbf{8.797} &	4.796 \\
    \bottomrule
    \end{tabular}
    \caption{The speed and count of reversion of the Ornstein-Uhlenbeck process. The closest values to the original cases are bold. The types differ in terms of (speed, volatility), with the parameters set to (1.0, 5.0), (1.0, 15.0), (2.5, 5.0), (2.5, 15.0), (5.0, 5.0), (5.0, 15.0).}
\end{table*}

\begin{table*} 
  \label{tab:jumpcountsize}
  \begin{tabular}{cccccccccccccccccc}
    \toprule
     &&&& \multicolumn{2}{c}{Jump Count} &&&&& \multicolumn{2}{c}{Jump Size} &&\\
    \midrule
    Length & Type & Original & MLP & MLP-CNN & LSTM & GRU & TCN & Original & MLP & MLP-CNN & LSTM & GRU & TCN \\
    \midrule
    \multirow{6}{*}{25} & 1	& 2.430	& 3.242	& \textbf{2.270}	& 1.499	& 1.816 & 1.498	& 0.012	& 0.016	& \textbf{0.011} & 0.008	& 0.009	& 0.008 \\
    & 2	& 2.886	& \textbf{2.827}	& 3.188	& 2.264	& 2.499	& 2.529	& 0.014 & \textbf{0.013} &	0.015 &	0.012 &	0.012 &	\textbf{0.013} \\
    & 3	& 2.394	& 1.613	& \textbf{2.246}	& 1.619	& 1.474	& 2.127	& 0.017 & 0.012 &	\textbf{0.015} &	0.011 &	0.012 &	\textbf{0.015} \\
    & 4	& 2.984	& \textbf{2.999}	& 2.336	& 2.093	& 2.849	& 2.857	& 0.021 & 0.020 &	0.017 &	0.016 &	\textbf{0.021} &	0.022 \\
    & 5	& 2.276	& 3.700	& \textbf{2.126}	& 1.686	& 2.123	& 2.897	& 0.023 & 0.034 &	0.019 &	0.019 &	\textbf{0.023} &	0.029 \\
    & 6	& 2.508	& 2.645	& 3.096	& \textbf{2.485}	& 1.949	& 2.587	& 0.027 & \textbf{0.027} & 0.028 & 0.028 &	0.022 &	0.025 \\
    \midrule
    \multirow{6}{*}{100} & 1 & 8.637 & 10.210 &	\textbf{7.356} &	5.708 & 5.201 &	5.624 &	0.010 &	0.022 &	\textbf{0.008} &	\textbf{0.008} &	0.007 & 0.007 \\
    & 2	& 9.791	& 8.720	& 7.980	& 6.504	& \textbf{9.831}	& 8.312	& 0.011 & 0.009 &	0.009 &	0.009 &	\textbf{0.012} &	0.010\\
    & 3	& 6.072	& 12.663 & 7.587 & \textbf{5.509} & 7.351 & 9.938 & 0.014 & 0.022 &	\textbf{0.014} &	0.011 & 0.010 & 0.016 \\
    & 4	& 6.751	& 9.961	& 8.629	& \textbf{7.209}	& 7.966	& 6.180	& 0.017 & \textbf{0.018} &	0.015 &	0.014 &	0.015 &	0.011 \\
    & 5	& 2.751	& 6.336	& 7.330	& \textbf{5.154}	& 9.256	& 7.731	& 0.019 & 0.015 &	\textbf{0.017} &	0.014 &	0.024 &	0.015 \\
    & 6	& 2.657	& 9.691	& 8.969	& 6.301	& \textbf{6.156}	& 6.582	& 0.022 & 0.024 &	\textbf{0.020} &	0.016 & 0.016 &	0.017 \\
    \bottomrule
    \end{tabular}
    \caption{The count and size of jumps of the Jump diffusion model. The closest values to the original cases in each type and length are highlighted in bold. The parameters settings of the Jump Diffusion are mean and std of jump size and detailed settings are as follows: (0.1, 0.02), (0.1, 0.03), (0.15, 0.03), (0.15, 0.05), (0.2, 0.04), (0.2, 0.06)}
\end{table*}

The results suggest that the GAN framework effectively learns both the distributions and features of Brownian Motion and Geometric Brownian Motion, which represent the randomness of stock behavior. However, the performance of the LSTM and TCN generators on the Ornstein-Uhlenbeck process and Jump Diffusion requires further analysis. The poorer performance of these generators on the Ornstein-Uhlenbeck process is particularly evident in the final value distribution in  \textbf{Figure 2}. This process’s reverting characteristics rely more on its past values than other features, and the final value strongly depends on the sequence of log returns. Although LSTM is a specialized neural network for sequential data, empirical studies have shown that it performs poorly on the log return data. Therefore, the absence of accurate sequential information may lead to generated final values that fall outside the range of the original data.

\textbf{Figure 3} shows the statistical properties of the log return of both original and generated data by all generators. The overall box plots of generated data are similar to those of the original data. However, in some cases, such as type 1 generated by the TCN generator, the length 100 cases are not generated as accurately as the original data, producing log return data with large maximum, minimum, and standard deviation values. Additionally, some outliers are detected in type 4 generated by the MLP and TCN generators.

The characteristics of the Geometric Brownian motion are shown in \textbf{Figure 4}. Similar to Brownian motion, the statistical properties estimate different parameter settings for Geometric Brownian motion. Although Geometric Brownian motion is also based on the randomness of its movement, the box plots of generated data are exceptionally similar to those of the original data. The training data for Geometric Brownian motion features strong drift in its up-phase trends, especially in types 3 and 5, and this characteristic is well captured in the generated data.

\textbf{Table 1} summarizes the results for the Ornstein-Uhlenbeck process, including the estimated reverting speed and the number of reverting cases. Among the various generator architectures, MLP and LSTM are the most effective at reproducing a reverting speed close to that of the original data, while TCN shows limited performance in learning this feature. However, the results of the LSTM generator are somewhat contradictory since its Jensen-Shannon divergence values are greater than those of other generator architectures.
It’s important to note that reverting speed is estimated only for the reverting case, without providing any information about the divergence or reversion of the time series. Therefore, the results of learning distributions do not precisely reflect the learning performance of each temporal structure. Although there is no clear tendency in learning reverting count between the network types, MLP and GRU perform best at generating time series with a reverting count closest to the original data. Interestingly, while TCN tends to produce reverting cases that occur by concentration, its reverting speed results are much smaller than those of the original data. Moreover, as the time series length increases, the error between the reverting speed of the original data and the generated time series also increases, highlighting the limitations of the GAN framework in learning inherent reverting features.

We present the results of the generated Jump Diffusion model data, comparing the jump count and size to the original data in \textbf{Table 2}. For shorter lengths, MLP and MLP-CNN architectures show the best performance in training the number of jump cases within one time series, while LSTM and GRU networks perform better for longer lengths. In terms of jump size, there is no clear trend in learning across the different generator architectures, but the MLP-CNN architecture produces the closest jump sizes to the original data.

\begin{figure*}[ht]
  \centering
  \includegraphics[width=17cm, height=4cm]{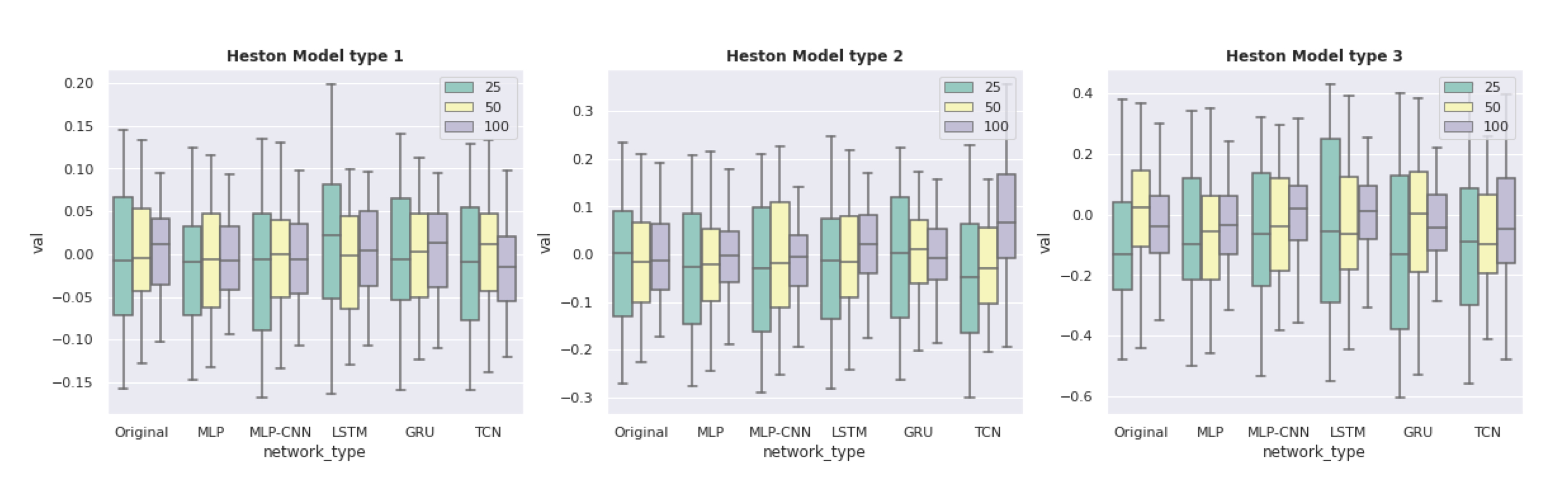}
  \caption{Box plots of 3 types of the Heston model. The first three box plots are statistical properties of original cases, and the following three boxes are the results of generated data corresponding networks. The types differ in terms of volatility of volatility, with the parameters set to be 0.01, 0.1, 1.0.}
\end{figure*}

\begin{figure*}[ht]
  \centering
  \begin{tabular}{cc}
       \includegraphics[width=8cm]{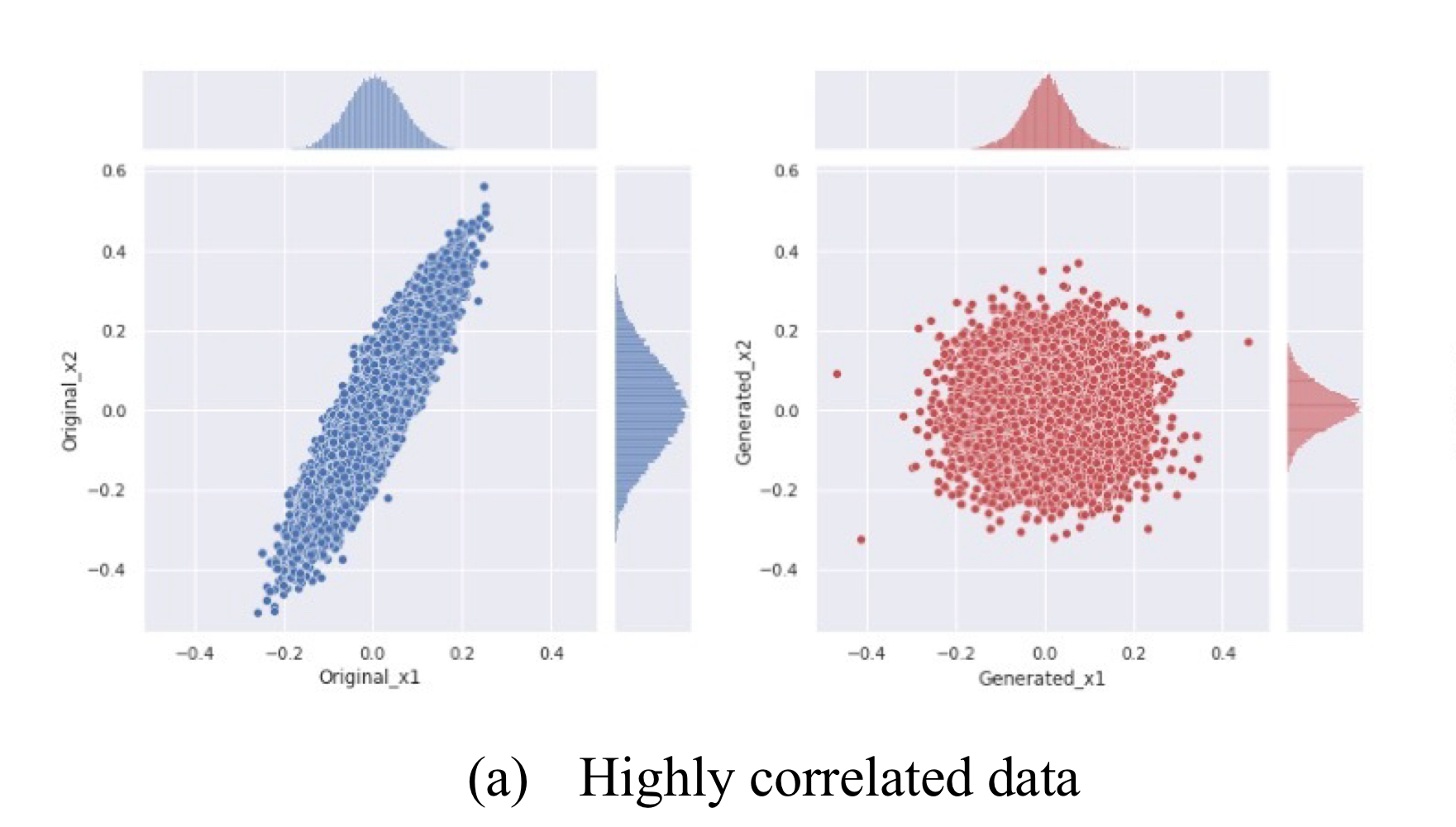}& \includegraphics[width=8cm]{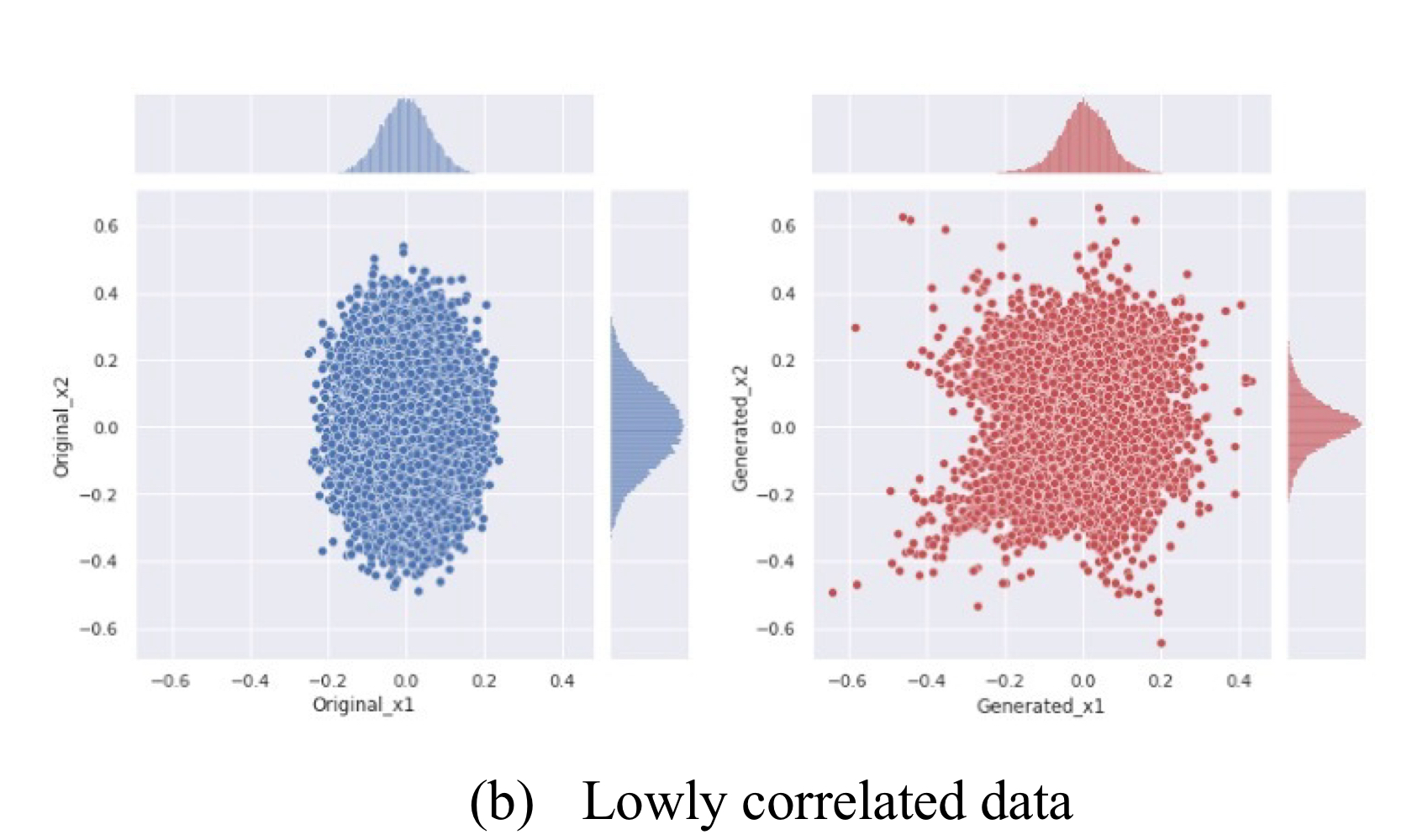} \\
  \end{tabular}
  \caption{Scatter plots of bivariate distribution and log return distribution of each variable. Blue plots refer to the original data, whereas red plots are generated data.}
\end{figure*}

The statistical properties of the volatility of volatility in the Heston model are estimated and described in the box plot in \textbf{Figure 5}. The volatility of volatility parameter settings for the Heston model are 0.01, 0.1, and 1.0 for types 1, 2, and 3, respectively. As the volatility of volatility increases, the box plots of the generated data deviate significantly from the box plot horizon of the corresponding length of original data. In Heston model type 1, each generator’s general performance shows remarkable results except for one case: the maximum and minimum values generated by the LSTM generator with length 25 are far from those of the original data. The box plots of type 2 are within the acceptable range, except for the TCN generator with a length of 100. For this length, all components of the box plot are shifted upwards, and the distribution of volatility of volatility is more platykurtic than the original data. In contrast to types 1 and 2, the performance in type 3 is slightly poorer, especially for LSTM and GRU generators with shorter lengths and the TCN generators with longer time series. Although the mean of LSTM and GRU cases with a length of 25 is similar to that of the original data, the standard deviation is significant. Additionally, the box of the TCN generator is relatively larger than those of the other networks and the original data.

\subsubsection{Multivariate time series generation}

We also conducted GAN training on multivariate Geometric Brownian motion. The generator and discriminator are composed of simple MLP layers and trained with the Binary Cross Entropy loss function. The evaluation for multivariate data is conducted from two perspectives: for each independent variable and for variable dependency. Firstly, the Jensen-Shannon divergence for each variable is estimated and presented in \textbf{Table 3}. 

\begin{table}
  \label{tab:jslogreturn}
  \begin{tabular}{ccc}
    \toprule
     & High correlated data & Lowly correlated data \\
    \midrule
    x1 & 0.0702 & 0.0874 \\
    x2 & 0.306 & .185 \\
    \bottomrule
    \end{tabular}
    \caption{The Jensen-Shannon divergence of the log return distribution.}
\end{table}

In the context of multivariate data learning, the performance of the plain GAN is limited in learning each variable and variable dependency, as shown in \textbf{Figure 6}. The generated distribution shapes of x1 and x2 do not resemble those of the original distribution. Additionally, as summarized in \textbf{Table 3}, the Jensen-Shannon divergence for variables x1 and x2 between the original and generated data is also imbalanced, indicating that the GAN tends to generate data dependent on only one variable, x1, which has smaller volatility than x2. One possible reason for this is that the discriminator does not separately distinguish between the two variables but instead provides feedback on both variables together. Consequently, the information on x2 with larger volatility may have been offset.
The bivariate distribution between x1 and x2 is also very different from the original. While the high correlation is well described in the original bivariate distribution that extends in one direction, the generated bivariate distribution forms a circular shape. In the low correlation setting, the bivariate distribution of the generated multivariate data is randomly spread, unlike the original, more concentrated data. The essential variable dependency feature, the correlation between x1 and x2, is also compared. For highly correlated data, the correlation between x1 and x2 is 0.8993 for the original data and 0.8249 for the generated data. Meanwhile, for the low correlation setting, the correlation between x1 and x2 is 0.001062 for the original data and -0.05633 for the generated data. Although the generated correlations are not exactly the same as the original correlations, the differences remain relatively low, 0.0744 and 0.057395, for high and low correlation cases, respectively. These results indicate that while the GAN can capture some aspects of the correlations between variables, it struggles to accurately reproduce the individual distributions and dependencies in the multivariate setting. This limitation suggests that further refinement of the GAN architecture or the training process may be necessary to better model multivariate financial time series.

\section{Conclusion}
In this paper, a vanilla GAN framework is investigated for its capability to learn significant temporal structures in financial time series rather than solely evaluating the GAN’s learning performance. We have concluded several findings by conducting extensive experiments with various network architectures on five stochastic processes representing different temporal structures.
In univariate data generation, the GAN models generally capture the different temporal structures but struggle with detailed features. The performance of GAN in learning the distributions and temporal structures of Brownian Motion and Geometric Brownian Motion is superior, yet some generator architectures, particularly the TCN-based generator, struggle to learn the reverting speed of the Ornstein-Uhlenbeck process. Additionally, only MLP and MLP-CNN-based generators can capture the two-hump-shaped distributions, whereas others create right-skewed distributions. For the Heston Model, the distributions of volatility of volatility are sometimes more platykurtic than the original data, and this trend worsens as the volatility of volatility increases. When it comes to multivariate data generation, the vanilla GAN framework is very limited in capturing the dependencies between multivariate time series. The generated distributions of x1 and x2 do not resemble the original distributions and look alike, indicating that the GAN tends to generate data dependent on only one variable, x1, which has smaller volatility than x2.

The cause of the limited replication of detailed features of the temporal structure remains for further study. It is premature to conclude that the adversarial framework of GAN is ineffective, as many other potential factors could affect the results, such as superficially constructed networks or sequence length. Although the GAN models in this experiment do not include complex neural network architectures, incorporating additional architectures, such as attention mechanisms, may result in better performance in learning time series dependencies.
In summary, GAN models have shown potential as financial simulators without mathematical assumptions, but the detailed replication of temporal structures or multivariate time series still requires additional attention.

\bibliographystyle{ACM-Reference-Format}
\bibliography{reference}
\end{document}